\documentclass{sig-alternate-05-2015}

\usepackage[numbers,sort&compress]{natbib}

\begin{document}

\setcopyright{acmcopyright}



\conferenceinfo{epiDAMIK: Epidemiology meets Data Mining and Knowledge discovery}{August 5, 2019, Anchorage Alaska, USA}


%

\title{Machine learning in healthcare - a system's perspective}
\subtitle{[Position paper]}
%
%
%
%
%

\numberofauthors{2} 
%
\author{
%
%
\alignauthor
Awais Ashfaq\\
       \affaddr{Center of Applied Intelligent Systems Research, Halmstad University. Sweden}\\
       \affaddr{Halland Hospital, Region Halland. Sweden}\\
       \email{awais.ashfaq@hh.se}
       \alignauthor
Slawomir Nowaczyk\\
       \affaddr{Center of Applied Intelligent Systems Research, Halmstad University. Sweden}\\       
       \email{slawomir.nowaczyk@hh.se}
       }


\maketitle
\begin{abstract}
A consequence of the fragmented and siloed healthcare landscape is that patient care (and data) is split along multitude of different facilities and computer systems and enabling interoperability between these systems is hard. The lack interoperability not only hinders continuity of care and burdens providers, but also hinders effective application of Machine Learning (ML) algorithms. Thus, most current ML algorithms, designed to understand patient care and facilitate clinical decision-support, are trained on limited datasets. This approach is analogous to the Newtonian paradigm of \textit{Reductionism} in which a system is broken down into elementary components and a description of the whole is formed by  understanding those components individually. A key limitation of the reductionist approach is that it ignores the component-component interactions and dynamics within the system which are often of prime significance in understanding the overall behaviour of complex adaptive systems (CAS). Healthcare is a CAS.

Though the application of ML on health data have shown incremental improvements for clinical decision support, ML has a much a broader potential to restructure care delivery as a whole and maximize care value. However, this ML potential remains largely untapped: primarily due to functional limitations of Electronic Health Records (EHR) and the inability to see the healthcare system as a whole. This viewpoint (i) articulates the healthcare as a complex system which has a biological and an organizational perspective, (ii) motivates with examples, the need of a system's approach when addressing healthcare challenges via ML and, (iii) emphasizes to unleash EHR functionality - while duly respecting all ethical and legal concerns - to reap full benefits of ML.
\end{abstract}

%
%


%
%

%
%



\section*{Introduction}
System's thinking is a holistic approach to understand Complex Adaptive Systems (CAS) by not only focusing on individual components of the system but also the interconnections between those components \cite{wolfram2002new}. A CAS is (i) a collection of several individual agents, (ii) without centralised co-ordination, (iii) with freedom to act in ways that are not always totally predictable and, (iv) whose actions are interconnected so that one agent's actions change the context for other agents resulting in novel characteristics exhibited by the system as a whole. Put differently, the overall output of a CAS is not equal to the sum of outputs of all its sub-systems. Examples include, but not limited to, the global climate, the financial market and the healthcare system.

Since the beginning of the 21st century, healthcare has been widely framed and studied as CAS on many levels such as diseases, patients, epidemiology, care-teams, hierarchy, practices and education to unravel the complexity and improve decision-making \cite{dammann2014systems,plsek2001challenge,plsek2001complexity,earn2000simple}. The complexity triggers in healthcare can be broadly described from a biological (\textit{intra-human} e.g. the immune system) and organizational (\textit{inter-human} e.g. multidisciplinary care teams) perspective.

From a biological perspective, an important source of complexity is the uncertainty in medical knowledge: explained by the complex human biology which is subject to nearly constant change within physiological pathways due to a complex series of gene/environment interactions \cite{densen2011challenges}. Concurrently, International Classification of Diseases (ICD-10) specified over 68,000 diagnoses and the list keeps growing as we await ICD-11. In order to cure or alleviate patient sufferings, clinicians practice thousands of drugs and therapies. As a result, the gap between the rapidly advancing medical knowledge base and the application of that knowledge to escalating population size continues to widen \cite{world2015world}. Since clinical decisions deal with human life, we want to be as certain as possible about practice and desired clinical outcomes. Put differently, unravelling biological complexity aims at reducing the uncertainty in clinical decisions.
 
From an organizational perspective, an important source of complexity is healthcare structure, which - unlike an engineered complex system - is a system that emerged over time with independent actors (patients, care-providers, technologists, tax-payers; and hospitals, clinics, laboratories, government etc.) involved \cite{rouse2008health}. The actors often have distinct leaderships, budgets, goals, regulations and tools of operation. Controlling the output of such a socio-technical complex system is - if not impossible - very challenging because of the high degree of inter-relatedness among the roles of actors which renders the overall system output not equal to the sum of outputs of all the sub-systems. Put differently, unravelling organizational complexity aims at identifying problem areas in healthcare where interventions can have significant impact on the overall system output.

\section*{Healthcare complexity and Machine Learning}
The application of ML in healthcare is widely anticipated as a key step towards improving care quality and curbing care costs \cite{hinton2018deep}. A boon to this anticipation is the widespread adoption of Electronic Health Records (EHRs) in the health system. In Sweden, EHRs were introduced in the 1990s and by 2010, 97\% of hospitals, and 100\% of primary care doctors used them for their practice \cite{adamski2014overview}. EHRs are real-time digital patient-centred records that allow secure access to authorized care-providers across multiple healthcare centres when required. This digitization in healthcare generates unprecedented amounts of clinical data, which when coupled with modern ML tools provides an opportunity to expand the evidence base of medicine and facilitate decision process. 

For instance, inpatient crowding (or high levels of hospital occupancy) are often associated with reduced quality of care and access burden on care-providers \cite{weissman2007hospital}. The benefits of discharge are both monetary: hospital stays are expensive, bed availability is increased; and non-monetary: patients are less prone to medical complications and can spend more time with family and return to work. However the benefits depend on patient outcome (prognosis), so discharging a patient is worthy only if he or she does not need to be re-admitted in near future, which makes it a prediction problem. Put differently, the ML predictive challenge is: which patients are likely to be readmitted using data available at the point of discharge.

Similarly, for instance, long waiting times for patients to access the next step in the care-process has long been a cause of dissatisfaction for patients and care-providers \cite{rechel2016public}. The benefits of reducing waiting times are several: better value for patient’s time, less work stress among care-providers, improved care quality and quick monetary reimbursements. However to reduce waiting times, we want to predict demand - or patient path - in order to allocate resources (staff, drugs, equipment etc.) efficiently. Put differently, the ML predictive challenge is: what path will the patient follow in the healthcare system using data available at the time of entry.

In the current era, where algorithms can beat GO champions \cite{gibney2016google} or drive a man to the hospital \cite{reed2018umb}, it is tempting to believe that the aforementioned ML challenges are not far-fetched. However, it is worth remembering that the utility of ML tools largely hinges on underlying data and in most situations, access to complete healthcare records is extremely challenging. Though there are inklings of AI in medicine, the necessary resources - data - are still lacking. The Dataset Information Resource\footnote{\url {https://cci-hit.uncc.edu/dir/index.php/Welcome_to_DIR}} (DIR) \cite{shi2018developing} describes over 12 commonly used EHR datasets for research projects, of which only 2 are publicly available \cite{johnson2016mimic,NHANES}. The more comprehensive datasets with 100,000+ subjects and integrating healthcare information from diverse care points are proprietary and often require approvals from one or more advisory committees along with access charges \cite{mscan,wolf2019data,gunaseelan2019databases,THIN,PREMIER}.

Advancements in medical knowledge and guidelines have progressed a paradigm shift in healthcare: from care in a single unit to care across multiple units with varying but specialized expertise. It is often referred as fragmentation in medicine. Thus, patient care (and of course data) is split along multitude of different facilities and computer systems and integration of this information into a single system faces numerous challenges, primarily from an organizational perspective \cite{oyeyemi2018interoperability}. These include privacy and security concerns, lack of acceptable standardized data formats, use of proprietary technologies by disparate vendors, costly interface fees and more. As a result, care-providers are impeded from accessing complete datasets and thus unable to understand all aspects of the patient health journey.

Just as humans are better equipped to understand the world when given complete facts, so too are algorithms. As a consequence of the fragmented and siloed landscape of healthcare, the potential of ML algorithms to understand care patterns remains largely untapped, both due to \textit{unavailability} and \textit{inaccessibility} of necessary data. For instance, one barrier to prediction studies (such as in the readmission example) is that patient information after the course of treatment - the true outcome data - is not always available in EHR. Though, patient reported outcome and experience measures are being developed and validated, their integration into EHRs is scarce \cite{gold2018implementation}. Data from health devices at homes – that monitor patient’s health beyond the hospital radar - are also an important source of information, yet their integration to existing EHRs is a challenge \cite{genes2018smartphone}.

Similarly, one hurdle to predict patient flows is the lack of interoperability between EHRs in different care chains. In healthcare, we consider care fragments (primary, secondary, emergency etc.) as independent bodies; however, they constitute a single body from a patient's perspective who travels through them during the care process. Thus addressing the ML predictive challenges would require accessing and merging data from all levels of the care chain to have a holistic (complete) approach to healthcare delivery. The urge of a holistic approach is highly emphasized today; else - given the huge number of ML applications in healthcare - we might soon face another wave of interoperability challenges. Only this time, it will be between ML prediction models for different care fragments.
\section*{Text focus and context demise}
The process of training most current ML algorithms on limited and often different health datasets, to understand patient care and facilitate decision-support is, to a large extent, analogous to the Newtonian paradigm of a \textit{clockwork universe} that aims to understand a system by breaking it down into elementary components and understanding those components to form a description of the whole. This mode of scientific inquiry is often referred as Reductionism (as opposed to Holism) following the belief that any system can be explained by analysing the most basic components of the system. Reductionism has guided scientific reasoning for centuries with great success such as the development of the cell theory or formulation of the periodic table etc. that are fundamentals of explaining a wide variety of things that we encounter in our lives. Despite being a very powerful approach in explaining individual components, Reductionism struggles to understand the new emergent behaviour that results due to non-linear interactions between the individual components. In the context of ML on limited EHR data, one might consider a ML model trained to predict hospital admission given Emergency Department (ED) data. The developed model might exhibit strong discrimination ability and rightfully recommend transfer of a sicker patient from ED to the hospital. However, the utility of this decision largely hinges on the availability of resources in the hospital (beds, care-providers) at the time of admission and transferring a patient without assuring resource availability in the hospital might deteriorate the health state of an already sick patient.
\section*{Conclusion}
EHRs are, in sum, valuable resources for clinical and organizational decision-making in healthcare. However, in order to reap full benefits of ML in healthcare, we need to realize the limitations of existing datasets and appreciate a system's approach to model the complex healthcare landscape. In the context of healthcare data, a system's perspective would mean a comprehensive data resource covering complete clinical, operational and financial information of care processes at individual, organizational and population levels. The Institute of Medicine also recommends including  social and behavior measures into patient EHRs (stress, isolation levels, physical activity, geocoding etc.) to better characterize the diverse range of factors that drive complex diseases and influence individual and population health \cite{institute2014capturing}. The challenge of interoperability among different EHR systems is, albeit hard but, not insurmountable \cite{wolf2019data,10.1093/ije/dyz262}. Simultaneously, the rejiggering of EHRs would demand novel research challenges in fields of data security and differential privacy to create scalable and secure data warehouses to store sensitive medical information and facilitate responsible access to researchers when required. Dual-purposing of EHRs also demands reorientation of ethical and legal rules because in addition to medical professionals, EHRs are being accessed by data scientists and administrators. 
\bibliographystyle{unsrtnat}
\bibliography{sigproc}
\end{document}